\documentclass[aps,twocolumn,prl,superscriptaddress,longbibliography,notitlepage]{revtex4-2}
\usepackage[T1]{fontenc}
\usepackage{amsmath}
\usepackage{amssymb}
\usepackage{graphicx}
\usepackage{glossaries}
\usepackage{bm}
\graphicspath{{figures/}}

\makeatletter

\usepackage{xcolor}
\definecolor{darkblue}{rgb}{0.1,0.2,0.6} 
\definecolor{lightblue}{rgb}{0.1,0.1,1.0}
\definecolor{darkred}{rgb}{0.8,0.1,0.2}
\usepackage{float}
\newcommand{\mbf}[1]{\mathbf{#1}}
\newcommand{\mrm}[1]{\mathrm{#1}}

\makeatother

\usepackage{babel}
\babelprovide[hyphenrules=english]{en}

\begin{document}
	\global\long\def\bra#1{\left\langle #1\right|}
	\global\long\def\ket#1{\left|#1\right\rangle }
	
	\title{Apparent bistability from weak long-range interactions}
	
	\author{Achilleas Lazarides}
	\affiliation{Loughborough University, Loughborough, Leicestershire LE11 3TU, UK}
	\author{Andrea Pizzi}
	\affiliation{Cavendish Laboratory, University of Cambridge, Cambridge CB3 0HE, United Kingdom}
	\begin{abstract}
		Bistability, or the coexistence of two stable phases, can be broken by a bias field $h$ destabilising one of the phases via the nucleation and growth of defects. Strong long-range interactions, $1/r^\alpha$ with $\alpha$ less than the system's dimensionality $d$, can suppress the proliferation of defects and restore bistability. The case of weak long-range interactions $d<\alpha < d+1$ remains instead poorly understood. Here, we show that it supports \emph{apparent} bistability: While the system has in principle a unique stable phase, it appears bistable for all practical purposes for $\alpha < \alpha_c$, with $\alpha_c > d$ behaving like a genuine critical point. At the core of this is an exponential scaling of the critical droplet size $R_c\sim h^{-1/(\alpha - d)}$ with $\alpha$, which makes nucleating destabilizing droplets extremely unlikely for $\alpha < \alpha_c$, and such that $\alpha_c$ is mostly independent of system size.  In support of these conclusions we provide  field-theoretical arguments and numerics on a probabilistic cellular automaton. Overall, our results  offer a way to rethink phase stability in systems with long-range interactions as well as a new route to achieve practical bistability.
	\end{abstract}
	\maketitle
	
	Many physical systems can settle into one of two stable states, a property known as bistability~\cite{binder1987theory}. This underlies familiar phenomena such as the coexistence of ice and water, the two magnetic orientations in a ferromagnet, filling and wetting phenomena at surfaces~\cite{Bonn2009}, and the triple phase coexistence of ouzo~\cite{sibley_coexisting_2024}. Often, however, a small external bias field or perturbation renders one of the two states less favorable. The system may remain in such a metastable state for a long time, but will eventually transition to the more stable one~\cite{langer1969statistical}.
	
\begin{figure}[h]
		\centering
		\includegraphics[width=\linewidth]{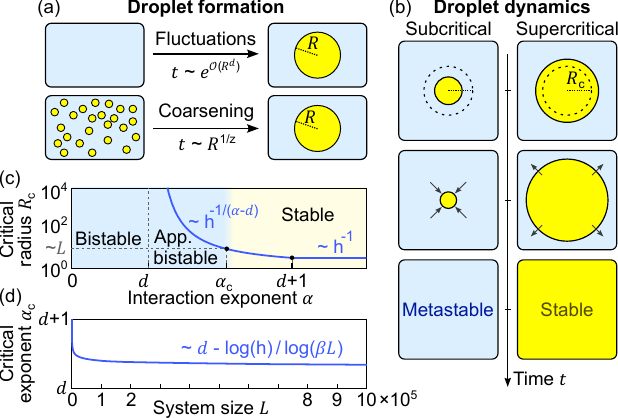}
		\caption{\textbf{Phenomenology of apparent bistability.}
			(a) Droplets of radius $R$ can nucleate from a clean metastable phase due to fluctuations, a slow process requiring exponentially long times $t \sim e^{\mathcal{O}(R^d)}$, or from an initial mixture of phases via coarsening dynamics, in a shorter time $t \sim R^{z}$.
			(b) Once nucleated, subcritical droplets shrink and disappear, while supercritical droplets grow.
			(c) The critical droplet radius $R_c$ scales with the bias field $h$ as $h^{-1}$ in the short-range regime where interactions decay with power $\alpha > d+1$ and as $R_c \sim h^{- 1/(\alpha - d)}$ in the weak long-range regime $d < \alpha < d + 1$, diverging rapidly for $\alpha \to d^+$.
			(d) The critical radius $R_c$ becomes comparable to the system size $L$ at $\alpha = \alpha_c$. For $d < \alpha < \alpha_c$, nucleation of a supercritical droplet is extremely unlikely and the system apparently bistable. The exponential divergence of $R_c$ with $\alpha$ makes $\alpha_c$ depend weakly on $L$ and behave like a genuine transition point.}
		\label{Fig1}
	\end{figure}

	This transition is usually understood in terms of droplet theory~\cite{fisher1967theory, langer1967theory, langer1969statistical, gunther1980goldstone, tomita1992statistical,rikvold1994metastable}. Droplets of the stable phase can spontaneously nucleate out of a metastable sea due to fluctuations, or form via coarsening from an initial mixture of the two phases~\cite{bray_domain-growth_1993,Bray:2010cb}, Fig.~\ref{Fig1}(a). The former takes an exponentially long time $t \sim e^{\mathcal{O}(R^d)}$, with $R$ the droplet radius and $d$ the dimensionality, whereas the latter only takes $t \sim R^{z}$ with $z$ a dynamical exponent. Small droplets then shrink due to surface tension and disappear, whereas large droplets grow until the entire system is stable, Fig.~\ref{Fig1}(b). A critical droplet radius $R_c$ separating these scenarios stems from the competition between bulk and boundary droplet energies. For example, in a two-dimensional Ising model at low temperature, $H = - J \sum_{\langle i,j \rangle} \sigma_i \sigma_j - h \sum_i \sigma_i$ with $J > 0$, a small field $h$ such that $|h|<J$ destabilizes one of the two ferromagnetic phases, the bulk and surface energies of a droplet read $\sim \pi R^2 |h|$ and $\sim 2 \pi R J$, respectively, and balancing them yields $R_c \sim J/|h|$~\cite{rikvold1994metastable}.
	
	The idea of surface tension at the core of standard droplet theory assumes short-range interactions. If the interactions are strong long-range, $1/r^\alpha$ with $\alpha < d$, the droplet picture breaks and the critical droplet size $R_c$ can diverge with system size, restoring bistability in the thermodynamic limit. This scenario has long been studied for classical systems~\cite{griffiths1966relaxation, penrose1971rigorous, chandra1989nucleation, antoni1995clustering, yamaguchi2004stability, mukamel2005breaking, campa2009statistical, gabrielli2010quasistationary, nishino2011macroscopic, gupta2017world}, and more recently for quantum ones~\cite{kastner2011diverging, defenu2021metastability, defenu2023long, defenu2024out, mattes2025long}, of relevance to Rydberg atoms, dipolar gases, and trapped ions among others~\cite{defenu2023long}. Excluding cases that ``piggyback'' on a symmetry~\footnote{For instance, for the two-dimensional Ising model, the critical droplet radius $R_c$ and metastable lifetime $\tau$ diverge for $h \to 0$. The existence of more stable phases in this case is unsurprising, indeed unavoidable given the $Z_2$ symmetry for $h = 0$. By contrast, we want here to focus on bistability in spite of a bias field $h$, that is, not directly relying on an underlying symmetry.}, bistability can also be restored for irreversible dynamics~\cite{bennett_role_1985}, as famously realized in Toom's rule~\cite{Toom1980,Grinstein2004} and as recently exploited in connection to time crystals~\cite{lazarides_time_2020,Gambetta2019,gambetta2019discrete,pizzi_bistability_2021,McGinley2022,machado_absolutely_2023}. Much less is instead known about the regime $\alpha > d$, for which the droplet picture should apply but can be substantially modified. In dimension one, it was shown for an Ising model that $R_c \sim h^{- 1/(\alpha - 1)}$ for $1<\alpha<2$~\cite{mccraw1980metastability,van2019nucleation}. In dimension two, numerics shows that $R_c$ diverges for $\alpha \to 2^+$~\cite{horikx2020metastability}, and metastability was investigated for long-range biaxial interactions~\cite{jacquier2024isoperimetric}. Despite this progress, a clear understanding of metastability and nucleation for $\alpha > d$ is missing~\cite{jacquier2025exploring}.
	
	Modern experimental realisations of this physics, often discussed in the context of  ``false vacuum decay'' in quantum simulators, include cold-atom and Rydberg platforms~\cite{Zhu2024ProbingFalseVacuum,gonzalezcuadra2025stringbreaking,xiang2025realtime} and quantum annealers~\cite{vodeb2025stirring} as well as gate-based quantum computers~\cite{farrell2024scalable,zemlevskiy2025scattering}. Rydberg platforms~\cite{gonzalezcuadra2025stringbreaking,xiang2025realtime}, in particular, naturally include power law interactions.
	
	Here, we investigate metastability in systems with power-law interactions. We develop a field-theoretical description of coarsening and droplet formation, predicting that in the weak long-range regime, $d < \alpha < d + 1$, the critical droplet size scales as $R_c \sim h^{- 1/(\alpha - d)}$. Such scaling has a remarkable consequence: Even for moderately large values of the bias $h$, the system can be \emph{apparently bistable}, namely appear bistable for all practical purposes, while strictly admitting a single stable phase. Namely, beyond $\alpha < d$ bistability extends in an apparent fashion to $d < \alpha < \alpha_c$, see Fig.~\ref{Fig1}(c). By numerically implementing a probabilistic cellular automaton, we show that the apparent phase boundary $\alpha_c$ displays the hallmarks of a \emph{bona fide} phase transition and is in practice indistinguishable from it.
	
	\textit{Field theoretical insights} ---
	We begin by elucidating the functional dependence of the critical droplet size $R_c$ on power-law exponent $\alpha$ and bias field $h$ using two complementary field theoretical approaches. The first assumes a Landau-Ginzburg-Wilson-type Hamiltonian for a scalar field $\phi(\mathbf{x})$ with power-law interactions~\cite{bray_domain-growth_1993}, $H[\phi]=H_{SR}+\int \mathrm{d}^d \mathbf{x} \int \mathrm{d}^d \mathbf{x}' \, \left( \phi(\mathbf{x}) - \phi(\mathbf{x}') \right)^2/|\mathbf{x} - \mathbf{x}'|^{\alpha}$. The short-range Hamiltonian $H_{SR}$ includes a gradient term suppressing spatial fluctuations and a potential $V(\phi)$ with two minima $\pm \phi_0$. For $\alpha>d$ and starting from a random field $\phi(\mathbf{x})$, the system relaxes via coarsening to one of the two uniform stable phases $\phi(\mbf{x})=\pm\phi_0$, up to thermal fluctuations. At time $t$ the system consists of intermixing domains of the two phases, of linear extent $\ell(t)$. The domain walls have an energy per unit area, and the energy minimization driven by $H_{SR}$ causes $\ell(t)$ to grow (coarsen) over time.
	
	Bray~\cite{bray_domain-growth_1993} showed, using power counting, that the energy per unit volume is $\rho_E\sim \ell^{-1}$ for $\alpha > d+1$ and $\rho_E\sim \ell^{-1/(\alpha-d)}$ for $d<\alpha<d+1$. Namely, in the short-range case the energetic cost of the walls is proportional to their density~\footnote{In a volume $\mathcal{L}^d$ there fit $N_\ell\sim(\mathcal{L}/\ell)^d$ objects of linear extent $\ell$, with total surface area $N_\ell \ell^{d-1}=\mathcal{L}^d/\ell$, so the wall area per unit volume is $\sim\ell^{-1}$.}, indicating they do not interact, whereas in the weak long-range case the energy is higher, implying interactions between walls. Switching on a bias field, $h\int \mathrm{d}^d\mathbf{x} \ \phi(\mathbf{x})$, makes one of the two phases metastable, penalizing it with energy density $\rho_h\sim h$. To find the $\ell_c$ above which the bias takes over from coarsening and the system relaxes to the stable phase, we balance the two contributions, $\rho_E\sim\rho_h$ and obtain $\ell_c\sim h^{-1}$ for $\alpha > d + 1$ and $\ell_c\sim h^{-1/(\alpha-d)}$ for $d < \alpha < d + 1$.
	
	This analysis can be modified for a single stable droplet, and yields a critical droplet size $R_c\sim h^{-1}$ for $\alpha > d+1$ and $R_c\sim h^{-1/(\alpha-d)}$ for $d < \alpha < d + 1$. To see this, we calculate the excess energy (over the energy of a uniform spin or field configuration) of a droplet of linear extent $R$ in a coarse-grained field theory with long-range interactions. The droplet could be circular or some other shape, but we assume that it is smooth and that the field $\phi$ is uniform inside the droplet and uniform outside of it. We use the coarse-grained Hamiltonian appearing in Ref.~\cite{bray_domain-growth_1993}, which should describe the universality class of Ising-type spin models like ours,
	\begin{equation*}
		\mathcal{H} = \int\, d \mathbf{x} \left[ \frac{1}{2} \left(\nabla \phi\right)^2 + V(\phi)\right] + V_{NL}\left[\phi\right],
	\end{equation*}
	with $\phi$ the coarse-grained spin field. The first integral is the local part of the Hamiltonian. The potential $V(\phi)$ has two minima at $\pm \phi_0$, imposing the Ising-like symmetry of the field. We take the second, non-local term to be~\cite{bray_domain-growth_1993}
	\begin{equation}
		V_{\mathrm{NL}}[\phi] = \int\,  d^d \mathbf{x} \int\,  d^d \mathbf{x}' \, \frac{\left[ \phi(\mathbf{x}) - \phi(\mathbf{x}') \right]^2}{|\mathbf{x} - \mathbf{x}'|^{d + \sigma}},
		\label{eq:VNL}
	\end{equation}
	namely a power-law ferromagnetic term resulting in an excess energy cost for anti-aligned fields. The energy is minimized by a uniform field $\phi(\mathbf{x}) = \pm \phi_0$.
	
	Let us now estimate the excess energy cost $\Delta E$ of a droplet of radius $R$, with field $+\phi_0$ inside and $-\phi_0$ outside. For the local part of the Hamiltonian, one uses standard techniques (see, e.g., Refs.~\cite{Schick1990,Dietrich1988}) to show that the transition between the two minima of $V(\phi)$ occurs locally, and that $\phi$ decays to $\pm \phi_0$ exponentially fast with the distance from the interface. Therefore the local part of the Hamiltonian contributes an excess energy proportional to the surface of the droplet, $\propto R^{d - 1}$. 
	
	The contribution of the non-local part of the Hamiltonian is more subtle. We treat $V_\mathrm{NL}$ as a perturbation to the local part of the Hamiltonian. The local part of the Hamiltonian forces a sharp interface between the two phases, and we thus evaluate $V_\mathrm{NL}$ for this interface. Assuming that it changes from $-\phi_0$ to $+\phi_0$ in a region of width much smaller than $R$, we approximate it by a step function and write
	\begin{equation*}
		E_\mathrm{NL} = V_\mathrm{NL}\left[\phi\right] = 8\phi_0^2\int_{|\mathbf{x}|<R}\, d^d \mathbf{x} \int_{|\mathbf{x}'|>R}\, d^d \mathbf{x}' \, \frac{1}{|\mathbf{x} - \mathbf{x}'|^{d + \sigma}},
	\end{equation*} 
	where we have separated the integral into four parts corresponding to $\mathrm{x}$ and $\mathrm{x}'$ being both inside, both outside, or straddling the interface, and using the fact that two of these cases vanish and the other two are equal. Defining $\mathbf{x}=\mathbf{y}R$ and $\mathbf{x}'=\mathbf{y}'R$,
	we rewrite
	\begin{equation*}
		E_\mathrm{NL} = 8\phi_0^2 R^{d - \sigma}\int_{|\mathbf{y}|<1} d^d y \int_{|\mathbf{y}'|>1} d^d y' \, \frac{1}{|\mathbf{y} - \mathbf{y}'|^{d + \sigma}}.
	\end{equation*}
	The integrals no longer depend on $R$, and $E_\mathrm{NL}$ depends on $R$ only via $R^{d-\sigma}$. The total energy cost is
	\begin{equation}
		\Delta E = \sigma R^{d-1} + \gamma R^{d-\sigma}
		\label{eq:DeltaE}
	\end{equation}
	with $\sigma$ and $\gamma$ constants independent of $R$. 
	
	In Eq.~\ref{eq:DeltaE}, the first term dominates and the excess energy is proportional to the surface area if $\sigma>1$. This applies to local Hamiltonians and shows that the physics is essentially local, despite the power law. If $0<\sigma<1$, instead, the second term dominates and implies an interaction between the droplet walls. Overall,
	\begin{equation}
		\Delta E \sim \begin{cases}
			R^{d - 1} & \quad \text{for} \ 1< \sigma, \\
			R^{d - \sigma} & \quad \text{for} \ 0 < \sigma < 1.
		\end{cases}
	\end{equation}
	
	Finally, to derive the critical droplet size $R_c$, we note that upon adding a bias $h$ coupling to the field via a term $h\int d^d\mathrm{x}\,\phi(x)$, the energy cost of the droplet becomes
	\begin{equation}
		\Delta E_h = \epsilon h R^d + \sigma R^{d - 1} + \gamma R^{d - \sigma} - h\phi_0R^d,
	\end{equation}
	where $\epsilon$ is another $R$-independent constant. The critical droplet size $R_c$ is then given by $\Delta E_h(R_c)=0$, yielding
	\begin{equation}
		R_c \sim \begin{cases}
			h^{- 1} & \quad \text{for} \ 1< \sigma, \\
			h^{-1/\sigma} & \quad \text{for} \ 0 < \sigma < 1.
		\end{cases}
	\end{equation}

	\textit{A long-range droplet theory} ---
	The approach above is powerful in its simplicity: Its predictions follow from simple scaling arguments. We complement it with an explicit droplet theory for $d<\alpha<d+1$. Consider a lattice of Ising variables $s_i = \pm 1$ with ferromagnetic interactions $J_{ij} = 1/(N_{\alpha} r_{ij}^{\alpha})$, with $r_{ij}$ the distance between particles $i$ and $j$, and $N_{\alpha} = \sum_j 1/r_{ij}^{\alpha}$ a Kac normalization. Due to the interaction, each particle is subject to an effective field $n_i = \sum_j J_{ij} s_j$ created by the others. Call $M = \frac{1}{N} \sum_{j = 1}^N s_j$ the average magnetization. For $h=0$ there are two stable phases, one with $M > 0$ and one with $M < 0$. A field $h>0$ breaks the symmetry and renders $M > 0$ stable, because $n_i$ and $h$ are both positive and ``reinforce'' each other, and $M < 0$ metastable.
	
\begin{figure}
		\includegraphics[width=\linewidth]{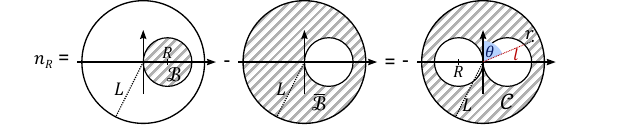}
		\caption{
			Graphical representation of Eq.~\eqref{eq. nR} to compute $n_R$, exemplified for $d = 2$. The shaded area represents the integration area and the integrand is $\left(N_\alpha |\bm{r}|^\alpha\right)^{-1}$.
			A circle of radius $R$ is denoted $\mathcal{B}$, and the area excluding two droplets tangent in the origin is called $\mathcal{C}$.
			\label{fig. mf-integrals}}
	\end{figure}

	The transition from metastable to stable is triggered by the nucleation and growth of defects. To locate it we work in a long-wavelength regime and treat space as continuous, which is justified because we target large critical droplet sizes. Consider a circular system of radius $L$, for $L \to \infty$. The Kac normalization reads 
	\begin{equation}
		N_\alpha = \int\, d \bm{r} \ \frac{1}{|\bm{r}|^\alpha} = \frac{\Gamma_d}{\alpha - d} \left(1 - \frac{1}{L^{\alpha - d}} \right),
	\end{equation}
	with $\Gamma_1 = 2$, $\Gamma_2 = 2 \pi$, and $\Gamma_3 = 4\pi$. In the integral above, to lift any unphysical self-interaction and the divergences at $|\bm{r}| \to 0$, we have implicitly restricted the integral to $|\bm{r}| \ge 1$, as we will also do in the following. We place a stable $+1$ droplet $\mathcal{B}$ of radius $R$ within a metastable $-1$ background $\bar{\mathcal{B}}$. The droplet is positioned such that the origin $\bm{r} = 0$ is on its surface, on which the effective field reads
	\begin{equation}
		n_{R}
		= \frac{1}{N_\alpha}
		\left( \int_{\mathcal{B}}\, \frac{d \bm{r^\prime}}{|\bm{r}|^\alpha}
		- \int_{\bar{\mathcal{B}}}\, \frac{d \bm{r^\prime}}{|\bm{r}|^\alpha} \right)
		= - \frac{1}{N_\alpha}
		\int_{\mathcal{C}} \frac{d \bm{r^\prime}}{|\bm{r}|^\alpha},
		\label{eq. nR}
	\end{equation}
	where $\mathcal{C}$ denotes the complement of two equal droplets tangent to each other in $\bm{r} = 0$, as graphically represented in Fig.~\ref{fig. mf-integrals}.
	
	For $d = 1$,
	\begin{align}
		n_R
		= -\frac{2}{N_\alpha} \int_{2R}^L  \frac{dr}{r^{\alpha}},
		= \frac{1}{L^{\alpha - 1}} - \frac{1}{(2R)^{\alpha - 1}}.
		\label{eq. nR d = 1}
	\end{align}
	
	For $d = 2$, we call $l = 2R\sin \theta$ the length of the segment that connects two points of the circle, with $\theta$ the angle between the segment and the tangent to the circle in one of its ends, see Fig.~\ref{fig. mf-integrals}, and get
	\begin{align}
		n_R
		& =
		-\frac{4}{N_\alpha} \int_0^{\frac{\pi}{2}} d \theta \int_{l(\theta)}^L  \frac{dr}{r^{\alpha - 1}}, \\
		& =
		\frac{1}{L^{\alpha - 2}} -\frac{2}{\pi}
		\frac{1}{(2 R)^{\alpha - 2}}
		\int_{0}^{\frac{\pi}{2}} \frac{d \theta}{(\sin \theta)^{\alpha - 2}}, \\
		& = 
		\frac{1}{L^{\alpha - 2}}
		-\frac{1}{(2 R)^{\alpha - 2}}
		\frac{1}{\sqrt{\pi}} \frac{\Gamma(\frac{3 - \alpha}{2})}{\Gamma(\frac{4 - \alpha}{2})},
		\label{eq. nR d = 2}
	\end{align}
	where we ignored the fact that the integral should have been restricted to $l(\theta) > 1$, namely, to $\theta < \arcsin{1/2R}$, which leads to subleading corrections for $R \gg 1$.
	
	For $d = 3$, a similar procedure yields
	\begin{align}
		n_R
		& =
		-\frac{2}{N_\alpha} \int_0^{2\pi} d \phi \int_0^{\frac{\pi}{2}} d \theta \ \sin \theta \int_{l(\theta)}^L \frac{dr}{r^{\alpha - 2}}, \\
		& =
		\frac{1}{L^{\alpha - 3}}
		- \frac{1}{(2 R)^{\alpha - 3}}
		\int_{0}^{\frac{\pi}{2}} \frac{d \theta}{(\sin \theta)^{\alpha - 4}}, \\
		& =
		\frac{1}{L^{\alpha - 3}}
		-\frac{1}{(2 R)^{\alpha - 3}}
		\frac{\sqrt{\pi}}{2}
		\frac{\Gamma(\frac{5 - \alpha}{2})}{\Gamma(\frac{6 - \alpha}{2})}.
		\label{eq. nR d = 3}
	\end{align}
	
	Altogether, we thus get $n_R = - \left[f_d(\alpha)/R\right]^{\alpha - d}$,
	with $f_1(\alpha) = \frac{1}{2}$, $f_2(\alpha) = \frac{1}{2} \left[ \frac{\sqrt{\pi} \Gamma(\frac{4 - \alpha}{2})}{\Gamma(\frac{3 - \alpha}{2})} \right]^{- \frac{1}{\alpha -2}}$, and $f_3(\alpha) = \frac{1}{2}\left[ \frac{2 \Gamma(\frac{6 - \alpha}{2})}{\sqrt{\pi} \Gamma(\frac{5 - \alpha}{2})} \right]^{- \frac{1}{\alpha -3}}$. The functions $f_d(\alpha)$ are well behaved, and in particular converge for $\alpha \to d$ from above, namely, $f_2(2^+) = 1$ and $f_3(3^+) = \frac{e}{4}$.
	
	On the edge of the droplet, and due to its curvature, the negative field from the background prevails on the positive field from the droplet, yielding $n_R < 0$. Increasing $R$ decreases the curvature and $|n_R|$. The competition between effective field $n_R < 0$ and bias field $h > 0$ determines whether the droplet expands or shrinks. If the droplet is small, $n_R$ prevails over $h$ and the droplet shrinks, whereas, if the droplet is large, $h$ prevails over $n_R$ and the droplet grows. Exact integration yields $n_R = - \left[f_d(\alpha)/R\right]^{\alpha - d}$, and the critical size $R_c$ has $n_{R_c} = -h$, namely,
	\begin{equation}
		R_c = f_d(\alpha) h^{- 1/(\alpha - d)}.
		\label{eq. Rc}
	\end{equation}
	
	As common for nucleation and droplet formation at low temperatures~\cite{tomita1992statistical}, we have obtained the critical droplet size from purely energetic arguments. The role of a finite temperature $T$ can in principle be included considering the free energy $F=E-TS$, but because the entropy $S$ of an interface scales as the energy $E$ also does, namely, with the area $R^{d-1}$ [cf. Eq.~\eqref{eq:DeltaE}], then at low temperatures $F \approx E$, and pure energetics yields a $R_c$ that does not depend on $T$. As for the short-range case~\cite{tomita1992statistical}, what strongly depends on the temperature is instead the timescale over which critical droplets nucleate.
	
	Overall, the two approaches above indicate that for $\alpha > d+1$ the critical droplet scales as in a short-range model, $\sim h^{-1}$, whereas for $d < \alpha < d+1$ we get $R_c\sim h^{-1/(\alpha - d)}$, with an exponent that diverges for $\alpha\rightarrow d^+$. The change in scaling at $\alpha=d+1$ has far-reaching effects. 
	
	Consider for instance the formation of droplets via coarsening as in Fig.~\ref{Fig1}(a). An infinitely large system ($L = \infty$) hosts an infinite mass of stable pockets, and these can coalesce into arbitrarily large droplets: Those above $R_c$ will expand and make the system transition to the stable phase. For a finite system size $L$, however, the mass of stable pockets is finite, and the droplets generated via coarsening, say of size $R \approx \beta L$, can be smaller than the critical droplet radius $R_c$. From Eq.~\eqref{eq. Rc} we get that this happens for $\alpha < \alpha_c$, with $\alpha_c \approx d + \frac{\log(1/h)}{\log \beta L/ f_d}$. Even for moderately large $h$, weak long-range interactions $d < \alpha < d+1$ can yield substantial $R_c$ and a seemingly critical $\alpha_c$ that drifts logarithmically slowly with system size $L$, see Fig.~\ref{Fig1}(d). This opens the door to apparent bistability, which we now set out to demonstrate numerically.
	
	\textit{A long-range majority voting rule} ---
	To concretely test the above ideas we introduce a long-range majority voting rule on a two-dimensional $L \times L$ square lattice. That of $d=2$ is the standard setting for nucleation and metastability and provides a natural short-range limit to benchmark long-range effects. Sites $i$ correspond to binary variables $s_{i} = \pm 1$, interacting with each other via a local field
	\begin{equation}
		n_i=\frac{1}{N_\alpha}\sum_{j \neq i} \frac{s_j}{r_{ij}^\alpha},
		\label{eq:local-field}
	\end{equation}
	with periodic boundary conditions and $N_{\alpha} = \sum_j r_{ij}^{-\alpha}$ a Kac normalization. At each discrete time $t = 0,1,2,\dots$, the $i$-th bit is set to $+1$ with probability
	\begin{equation}
		p(n_i)=\frac{1}{2}\left[1+\tanh\left(\frac{n_i+h}{T}\right)\right],
		\label{eq:prob-up}
	\end{equation}
	and to $-1$ otherwise. The motivation for this specific model is twofold. On the one hand, it satisfies detailed balance, as does any synchronous probabilistic cellular automaton with $p_i=\frac{1}{2}\left[1+\tanh\left(\sum_j J_{ij}s_j+h\right)\right]$ and $J_{ij}=J_{ji}$~\cite{Grinstein1985-sm-pca,lebowitz-sm-pca}, unlike Ref.~\cite{pizzi_bistability_2021}. Indeed, metastability can be strongly affected by the breaking of detailed balance~\cite{Toom1980,bennett_role_1985,Grinstein2004}, and preserving it is important to distill the role of long-rangeness. On the other hand, detailed balance does not come at the cost of sequential updates. In a traditional Hamiltonian setting, such as a long-range Ising model with Metropolis or Glauber dynamics, detailed balance is fulfilled but spin flips should be proposed one by one, which is computationally demanding. By contrast, a majority voting rule as in Eq.~\eqref{eq:prob-up} allows parallel updates of all bits, facilitating access to the large size and time scales that long-range systems involve.
	
	In Eq.~\eqref{eq:prob-up}, $h$ acts as a biasing field, with $h>0$ favoring $s_i = +1$, while $T>0$ plays the role of temperature and controls the amount of noise in the dynamics, with $p = 1/2$ for $T = \infty$ and $p(n_i)=\mrm{H}(n_i+h)$ a step function for $T=0$. For $h=0$, the model is symmetric under $s \rightarrow -s$ and  $p \rightarrow 1 - p$, yielding two stable phases with opposite magnetization. For a finite $h$, say $h > 0$, the state with $M > 0$ is stable whereas the stability of the state with $M < 0$ depends on $\alpha$, $h$, $T$, and $L$.
	
	\begin{figure}
		\centering
		\includegraphics[width=\linewidth]{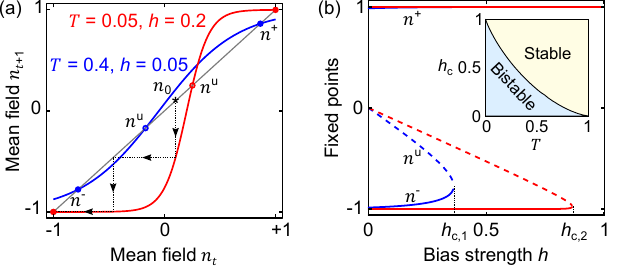}
		\caption{\textbf{Mean field analysis in the strong long-range limit $\bm{\alpha < d}$.}
			(a) Graphical fixed point analysis of the mean-field dynamics $n_{t+1} = 2 p(n_t) - 1$. The fixed points are the intercepts between $n$ (dashed) and $2 p(n) - 1$ (solid, for two sets of $T$ and $h$). The mid fixed point, $n^u$, is unstable: Starting from $n_0 < n^u$ yields $n_{\infty} = n^-$ (shown for the red line), whereas $n_{\infty} = n^+$ for $n_0 > n^u$.
			(b) Dependence of the fixed points on the bias $h$. For a strong bias, $h > h_c$, the system has a unique stable fixed point $n^+$. The transition happens at a field $h_c$, where $n^-$ and $n^u$ merge and annihilate. The dependence of $h_c$ on $T$ is shown in the inset.}
		\label{Fig2}
	\end{figure}
	
	\begin{figure*}
		\centering
		\includegraphics[width=\linewidth]{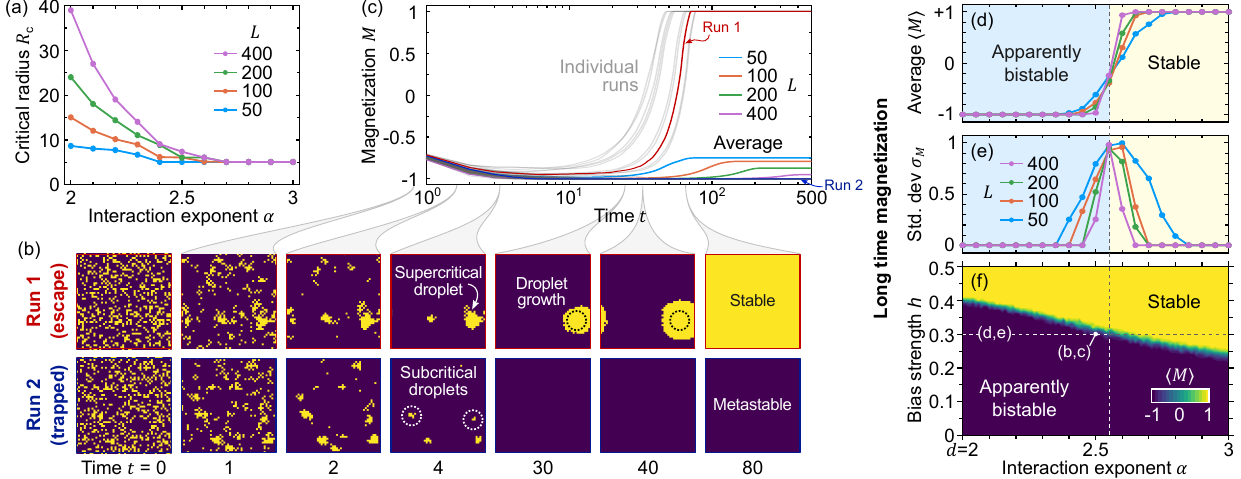}
		\caption{\textbf{Apparent bistability in a long-range majority voting rule.}
			(a) Critical droplet size $R_c$, obtained by preparing the system with a droplet, evolving it, and checking when its magnetization stays constant. The weak long-range interactions result in strong finite-size effects making $R_c$ scale with the considered system sizes $L$ for $d < \alpha < \alpha_c$.
			(b) Snapshot of two representative runs for $L=50$, starting from two random initial states with equal initial magnetisation. In the first, the short-time coarsening dynamics forms a supercritical stable droplet, that subsequently grows until engulfing the system. In the second, no such droplet is generated and the system falls back to the metastable state.
			(c) The dynamics of the magnetization is shown for many runs (light gray), all with the same parameters as (b). Some trajectories are similar to Run 1 in (b) (here marked in red), while others are similar to Run 2 (marked in dark blue). Increasing $L$ leads to larger $R_c$ and decreases the fraction of runs in which coarsening yields a supercritical droplet, flattening towards $-1$ the average magnetization over runs (colored filled lines).
			(d) Average magnetization $M$ at long times. For $\alpha > \alpha_c \approx 2.55$, the system reaches the stable phase ($M \approx 1$). For $\alpha < \alpha_c$, it instead stays in the metastable state ($M \approx -1$). The crossover sharpens for larger system sizes $L$, as does the peak of the standard deviation $\sigma_M$ over the runs (e), indicating an apparent phase transition.
			(f) A phase diagram is obtained plotting $\langle M \rangle$ in the ($\alpha,h$) plane. Here, $T = 0.1$, $h = 0.3$ in (a-e), and $\alpha = 2.5$ in (b,c).}
		\label{Fig3}
	\end{figure*}
	
	Intuition can be gained from the solvable limits $\alpha\rightarrow \infty$ and $\alpha\rightarrow 0$. The short-range limit, $\alpha\rightarrow\infty$, amounts to a nearest-neighbor majority voting rule and follows standard droplet theory: There is a finite $R_c \sim h^{-1}$ and supercritical droplets will eventually nucleate and lead the system to the unique stable phase. In the infinite-range limit, $\alpha = 0$, all sites are subject to the same mean field $n = L^{-2}\sum_i s_i$, which undergoes a discrete dynamics $n_{t+1}=2p(n_t)-1$ and can be studied with standard tools from dynamical system theory. Its fixed points solve $n = 2p(n)-1$ and are at most three, denoted $n^-$, $n^u$, and $n^+$. The unstable fixed point $n^u$ is the separatrix between the basins of attraction of the stable fixed points $n^-$ and $n^+$: If $n_0<n^u$ the system flows to $n^-$, while if $n_0>n^u$ it goes to $n^+$, see Fig.~\ref{Fig2}(a). For $h > h_c$ only a stable fixed point $n^+$ is present, that is, bistability is broken, see Fig.~\ref{Fig2}(b). The critical field $h_c$ decreases with temperature $T$ starting from $h_c=1$ at $T=0$ (inset). For $h < h_c$, a system with a stable droplet of size $R$ in a metastable background has $n_0 \sim (R/L)^2$, and $n_0 = n^u$ yields a critical droplet radius $R_c \propto L$, that is, the $R_c$ diverges in the thermodynamic limit and the system is bistable.
	
	The treatment for $\alpha = 0$ extends to the entire strong long-range regime $\alpha \le d=2$. This is because $1/N_\alpha$ in Eq.~\ref{eq:local-field} vanishes for $L \to \infty$ throughout this regime, so that flipping any finite number (not scaling with $L$) of spins on top of the metastable state does not affect $n_{i}$ -- the system is still only sensitive to one mean field parameter $n$ and the fixed-point analysis above applies~\cite{mattes2025long}.
	
	For the most interesting regime, $d < \alpha < d+1$, we rely on numerics. We expect no strict bistability: $R_c$ is finite and supercritical droplets will ultimately take the system to the stable phase. For \emph{accessible} time and length scales, however, we show that the system appears bistable for $d < \alpha < \alpha_c$, with $\alpha_c$ behaving like a genuine phase boundary. 
	
	We focus on $h > 0$, for which the stable phase has $M > 0$. First, we find the critical droplet size $R_c$ by initializing a stable droplet in a metastable background and monitoring the dynamics of the magnetization: Supercritical droplets grow, subcritical droplets shrink, and critical droplets yield $\mathrm{d}M/\mathrm{d}t = 0$~\footnote{Note that Fig.~\ref{Fig3}(a) assumes a single droplet placed in a square box of size $L$. With periodic boundary conditions, this is equivalent to a periodic array of droplets, namely, not a single droplet but rather a mixed set of both phases, not dissimilar from Fig.~\ref{Fig2}(b).}. This procedure yields a $R_c$ that, for small enough $\alpha$, grows with system size $L$, see Fig.~\ref{Fig3}(a). For $L \to \infty$ we in fact expect convergence to $R_c \sim h^{-1/(\alpha - d)}$, but the exponential divergence with $\alpha$ is such that $R_c \sim L$ for $\alpha < \alpha_c \sim d + \frac{\log \left(h^{-1}\right)}{\beta + \log L}$. As in Fig.~\ref{Fig1}(d), the logarithmic dependence makes the phase boundary $\alpha_c$ shift logarithmically slowly with $L$, leading to an apparent divergence of $R_c$ with $L$ for $d < \alpha < \alpha_c$.
	
	Inserting a single large stable droplet ``by hand'' was instrumental in obtaining $R_c$. Bistability is however more naturally understood in a scenario in which the system is initially metastable with a finite density of stable pockets. We thus prepare it with a negative initial magnetisation $M = -0.5$, so that the stable phase is initially in minority, and sit right at the left of $\alpha_c$. A short-time coarsening dynamics can nucleate one or more stable supercritical droplets, later leading to the stable phase, or not, letting the system fall back to the metastable phase [Fig.~\ref{Fig3}(b)]. Indeed, for $\alpha < \alpha_c$ increasing $L$ enlarges the effective critical droplet and thus the likelihood to fall back to the metastable phase, leading to apparent bistability [Fig.~\ref{Fig3}(c)]. For $\alpha > \alpha_c$, the critical droplet size $R_c$ is instead independent of $L$ and, because nucleation can happen at any point in the system, the probability that coarsening yields a supercritical droplet grows as $\sim L^2$. Together, these facts lead to the apparent transition at $\alpha_c$.
	
	The transition is better appreciated from the long-time magnetisation versus $\alpha$ in Fig.~\ref{Fig3}(d). This reveals a crossover between $M \approx -1$ (fall back to metastable) and $M \approx +1$ (reach the stable phase). The crossover sharpens for increasing system sizes $L$, consistently with a phase transition at $\alpha_c \approx 2.55$. This is reinforced by the fluctuations of the magnetization, shown in Fig.~\ref{Fig3}(e), which yield a peak at $\alpha_c$ that narrows for increasing $L$. Using these metrics, we obtain the two-dimensional phase diagram in the $(\alpha,h)$ plane shown in Fig.~\ref{Fig3}(f). The system is apparently bistable for a large region of the parameter space, and separated from the stable phase by what appears a genuine phase transition to standard diagnostics. In reality, for large system sizes and times the transition would disappear, but the large $R_c$ and slow drift of $\alpha_c$ with $L$ make it invisible for these system sizes. This is a direct consequence of the exponential divergence of $R_c$ with $\alpha$ as $\alpha\rightarrow d^+$, so that for \emph{any} $L$, no matter how large, there is an $\alpha_c$ below which the system appears bistable. By contrast, for the short-range regime $\alpha > d + 1$, the critical droplet size is independent of $\alpha$ and this behavior is not present. Note that the observed phenomenology is robust to noise, which is inherent in our stochastic model.

	\textit{Discussion and conclusion} ---
	We have studied bistability in systems with power-law interactions $1/r^\alpha$. We found that, in the presence of a bias field $h$, the lifetime of the metastable phase scales as $h^{-1}$ for $\alpha > d + 1$, and as $h^{-1/(\alpha -d)}$ for $d < \alpha < d+1$. The exponential divergence for $\alpha \to d^+$ underpins an apparent phase transition to bistability: The metastable state can be stable for all practical purposes not only for $\alpha < d$, but also for $d < \alpha < \alpha_c$. The enlarged boundary $\alpha_c$ appears a genuine transition point, as signaled by standard diagnostics and scaling analyses. The concept of apparent bistability highlights an important distinction between mathematical and physical stability, suggesting that even when true bistability is theoretically absent, weak long-range interactions can generate practically indistinguishable behavior. This offers a new route towards practical bistability and towards rethinking phase stability in a wide class of systems.
	
	The physics we have described is relevant for experiments with long-range interactions, such as trapped-ion quantum simulators~\cite{monroe2021programmable} and Rydberg atom arrays~\cite{saffman2010quantum,browaeys2020many}, in which one of two ground states is made unstable by a small bias field. As common with droplet-induced instabilities~\cite{rikvold1994metastable}, the spontaneous nucleation of a droplet can take astronomically large times and thus be hard to observe, both in numerics and in experiments. Instead, the stability of the droplets would be more easily studied, as in our numerics above, seeding a droplet into the system ``by hand'', and monitoring its dynamics (Fig.~\ref{Fig3}). For instance, the dynamics of seeded droplets has recently been studied in quantum simulators within the closely related context of false vacuum decay~\cite{manovitz2025quantum}.
	
	Future research should address apparent bistability in ``small-world'' systems with a few long-range~\cite{watts1998collective}, open systems with long-range jump operators~\cite{passarelli_dissipative_2022}, and long-range prethermal time crystals~\cite{machado2020long, pizzi2021classical, ye2021floquet, pizzi2021higher, jin2022prethermal,Yi-Thomas}, which indeed usually oscillate between two phases requiring a form of bistability. It would then be worth characterizing apparent bistability in the traditional Hamiltonian setting, in contrast to majority voting rules, for instance, in the Glauber dynamics of a long-range Ising model, although this would require a dynamics with asynchronous updates. Finally, and especially in light of the recent investigation of droplet nucleation in a quantum simulator~\cite{luo2025quantum}, open questions arise regarding apparent bistability in metastable quantum many-body systems~\cite{yin2025theory,lagnese2024detecting,Macieszczak2016}.
	
	\begin{acknowledgments}
		\textbf{Acknowledgments}. We thank Y.~Bar Lev, F.~Carollo, Robert L.~Jack, J.~Knolle, A.~Lerose, F.~Piazza, and H.~Zhao for discussions on this and related works. A.~P. acknowledges support by Trinity College Cambridge and A.~L.~ acknowledges support from the Leverhulme Trust Research Project Grant RPG-2025-063.
	\end{acknowledgments}

\end{document}